\documentclass[12pt]{iopart}

\usepackage{graphicx}% Include figure files
\usepackage{dcolumn}% Align table columns on decimal point
\usepackage{bm}% bold math

\expandafter\let\csname equation*\endcsname\relax

\expandafter\let\csname endequation*\endcsname\relax

\usepackage{color}

\usepackage{comment}
\usepackage{amssymb,amsmath,amsfonts,mathrsfs,amsthm,fancyhdr}
\usepackage{braket}
\usepackage[normalem]{ulem}
\usepackage{latexsym}
\def\qed{\hfill $\Box$}

\def\sH{\mathscr{H}}

\newtheorem{theorem}{Theorem}[section]
\newtheorem{lemma}{Lemma}[section]
\newtheorem{corollary}[theorem]{Corollary}

\begin{document}

\title{Survival probability of the Grover walk on the ladder graph}

\author{E. Segawa}
\address{Graduate School of Environment and Information Sciences, Yokohama National University, Hodogaya, Yokohama 240-8501, Japan}

\author{S. Koyama}
\address{Department of Applied Mathematics, Faculty of Engineering, Yokohama National University, 79-5 Tokiwadai, Hodogaya, Yokohama, 240-8501, Japan}

\author{N. Konno}
\address{Department of Applied Mathematics, Faculty of Engineering, Yokohama National University, 79-5 Tokiwadai, Hodogaya, Yokohama, 240-8501, Japan}

\author{M. Štefaňák}
\address{Department of Physics, Faculty of Nuclear Sciences and Physical Engineering, Czech Technical University in Prague, B\v
rehov\'a 7, 115 19 Praha 1 - Star\'e M\v{e}sto, Czech Republic }
\ead{martin.stefanak@fjfi.cvut.cz}

\begin{abstract}
We provide a detailed analysis of the survival probability of the Grover walk on the ladder graph with an absorbing sink. This model was discussed in Mare\v s et al., Phys. Rev. A {\bf 101}, 032113 (2020), as an example of counter-intuitive behaviour in quantum transport where it was found that the survival probability decreases with the length of the ladder $L$, despite the fact that the number of dark states increases. An orthonormal basis in the dark subspace is constructed, which allows us to derive a closed formula for the survival probability. It is shown that the course of the survival probability as a function of $L$ can change from increasing and converging exponentially quickly to decreasing and converging like $L^{-1}$ simply by attaching a loop to one of the corners of the ladder. The interplay between the initial state and the graph configuration is investigated.
%\keywords{Quantum walk \and Grover walk \and Survival probability \and Quantum transport}
\end{abstract}

\date{\today}

\maketitle

\section{Introduction}
It is well known that an irreducible random walk on a finite and connected graph hits arbitrary vertex in a finite time. 
Hence, if there is an absorbing sink in the graph, the survival probability of the random walk in the interior of the graph converges to zero in the long time limit. 
On the other hand, for quantum walks \cite{Aharonov1993,Meyer1996} the situation can be quite different, as pointed out by Krovi and Brun \cite{krovi2006a,Krovi:2006,krovi:2007} who found quantum walks with infinite hitting times. In general, the Grover walk \cite{Inui2004,Inui2005}, which is induced by a random walk \cite{HKSS}, survives in the long time limit even if there is a sink vertex. It is mathematically clarified in \cite{KonnoSegawaStefanak} that such a counter-intuitive property of the Grover walk is induced by (i) the homological structure in the graph which is characterized by the first Betti number; (ii) the pair of loops; (iii) the persistent eigenspace of the underlying random walk to the Dirichlet boundary condition on the sink. 

The non-zero survival probability is induced by the so called dark states or trapped states \cite{lukin:2000,poltl:2009,emary:2009,donarini_coherent_2019}. These are the eigenstates of the hamiltonian (in the continuous-time case) or the unitary evolution operator (in the discrete-time case) which are not affected by the sink. Dark states can significantly reduce the efficiency of quantum transport \cite{huelga:2013,hu:2018} or cause imperfect detection in quantum search \cite{thiel:2020a,thiel:2020b}. The suppression of dark states by noise \cite{Rebentrost_2009,caruso:2009,Chin_2010,le:2018,wertnik:2018}, symmetry breaking \cite{novo:2015,hu:2018b}, tailored initialization \cite{muelken:2012} or percolation \cite{stefanak:2016,mares2019,mares:2020b} was investigated.

Another counter-intuitive phenomena in quantum transport was identified in \cite{mares2020}, where it was found that on some graphs the transport efficiency can be significantly improved by increasing the distance between the source and the sink or by adding redundant branches to the graph. One such example, which we investigate in detail in the present paper, is the ladder graph of length $L$. The Grover walk is launched at one corner of the ladder graph and the sink vertex is set at its diagonal corner; see Fig.~\ref{fig:ladders}. Non-orthogonal but linearly independent basis vectors of the dark subspace, which prevent perfect transmission to the sink, can be constructed from face-cycles and paths connecting loops, see Figs.~\ref{fig:trapped_no_loops} and \ref{fig:trapped_loops}. One would expect that as the number of dark states increases with $L$ the transport efficiency should decrease and conversely, the survival probability should increase. However, the numerical simulations of \cite{mares2020} show the opposite behavior; the survival probability is monotone {\it decreasing} for sufficiently large $L$. This phenomenon might be called the ``Hashigo-sake\footnote{Hashigo-sake is a Japanese idiom which means barhopping; ``Hashigo" and ``sake" are ladder and beverage in Japanese.} phenomenon" of quantum walker. As discussed in \cite{mares2020}, the reason behind this unusual behaviour of the survival probability can be traced to a special structure of the dark subspace, in particular, one dark state with normalization which is decreasing with $L$. To clarify this argument, we consider various configurations of the ladder graph and show that the trend of the survival probability indeed changes when the specific dark state emerges. Moreover, we construct an orthonormal basis in the dark subspace with the Gram-Schmidt procedure. This allows us to derive explicit formulas for the survival probability as a function of the ladder length $L$ and the initial state for each considered configuration of the graph.

This paper is organized as follows. 
In Section~\ref{sec:notation}, we introduce the notation, define the Grover walk on several variants of the ladder graph with sink, and decompose the dark subspace following the formulation of \cite{KonnoSegawaStefanak}. 
In Section~\ref{sec:3}, the contribution of the face-cycles to the survival probability is computed. 
In Section~\ref{sec:4}, we obtain the contribution of the loops to the survival probability. We summarize our results in Section~\ref{sec:concl}. 
In the Appendix~\ref{app} we prove that there are no additional dark states described by special eigenstates of the transition matrix of a simple random walk.

\section{Notation}
\label{sec:notation}

\begin{figure}[t]
    \centering
    \includegraphics[width=0.8\textwidth]{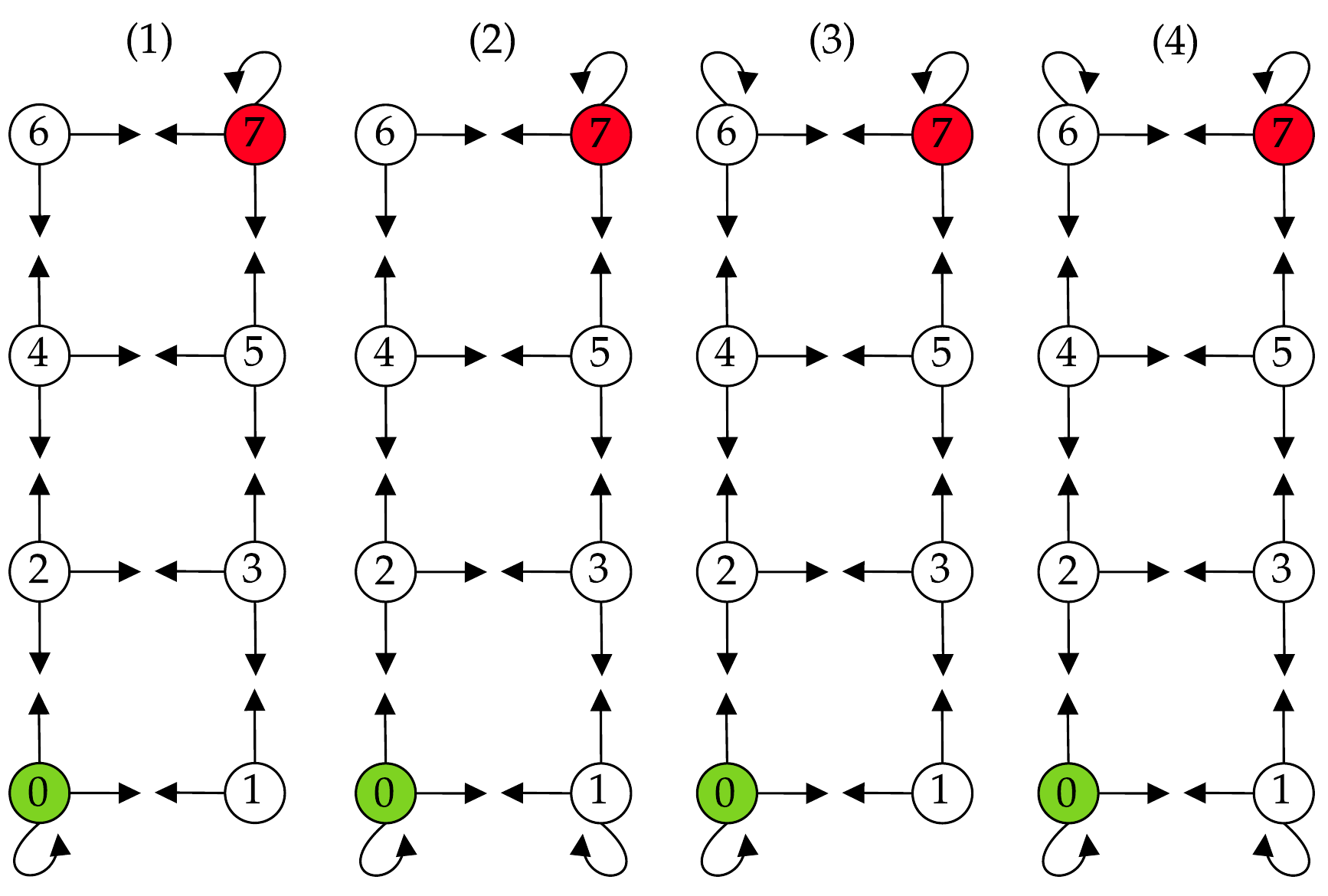}
    \caption{Four configurations of the ladder graph: (1) graph with loops only at the initial and sink corners, (2) additional loop at the bottom right corner, (3) additional loop at the top left corner, (4) graph with loops at all corner vertices. The excitation starts the Grover walk at the bottom left corner (vertex labeled as $0$, denoted by green). The sink is located at the top right corner (for the ladder of length $L=3$ it is the vertex $7$, denoted by red). }
    \label{fig:ladders}
\end{figure}

In the present paper we are interested in the survival probability of the quantum walk on the ladder graph, which comprises of two parallel mutually interconnected lines of finite length. We label the vertices in the left branch with even numbers $2j$ and in the right branch with odd numbers $2j+1$, with $j=0,\ldots L$. Here $L$ is the length of the ladder, i.e. the number of faces. The set of vertices of the ladder graph, which represents all possible walker's positions, is given by
$$
V = \left\{i|i=0,\ldots, 2L+1\right\}.
$$
The walker starts at the bottom left corner (vertex 0), and the absorption center (sink) is located at the top right corner (vertex $2L+1$). We investigate how is the survival probability influenced by adding loops to the remaining corners of the ladder (bottom right vertex 1 and top left vertex $2L$). For this reason we consider four variants of the ladder graph depicted in Figure~\ref{fig:ladders} : (1) graph with loops at the bottom left and the top right corners, (2) additional loop at the bottom right corner, (3) additional loop at the top left corner, (4) loops at all corner vertices. Note that the loop at the sink vertex is not relevant for the dynamics of the quantum walk, and we consider it for simplicity (the graph is 3-regular in the last configuration). In the first configuration of the ladder graph we can also omit the loop at the initial vertex, but we include it so that initial state is of the same form for all four settings.

We consider ladder as directed graph $G(V,E)$, where the edges encode directions which the walker will follow in the next step of the evolution. The set of edges can be decomposed into the set of arcs $E_a$ and the set of loops $E_l$, i.e. $E = E_a \cup E_l$. For every arc $e \in E_a$ there is an arc with opposite direction, which we denote by $\bar{e}\in E_a$. In our notation the set of arcs is given by
\begin{align*}
        E_a = & \left\{(2i,2i+1), (2i+1,2i)| i = 0,\ldots, L\right\}  \\
        & \cup \left\{(2i,2i+2)),(2i+1,2i+3)| i = 0,\ldots, L-1\right\}  \\
        & \cup \left\{(2i,2i-2),(2i+1,2i-1)| i = 1,\ldots, L\right\}.
    \end{align*}
The set of loops, according to the configuration of the graph, is given by
\begin{enumerate}
\item[(1)] $E_l =  \{(0,0), (2L+1,2L+1)\} ,$
\item[(2)] $E_l =  \{(0,0), (2L+1,2L+1), (1,1)\} ,$
\item[(3)] $E_l =  \{(0,0), (2L+1,2L+1), (2L,2L)\} ,$
\item[(4)] $E_l =  \{(0,0), (2L+1,2L+1), (1,1), (2L,2L)\}  .$
\end{enumerate}
Every directed edge $e=(v,w) \in E$ with $v,w \in V$, either arc or loop, corresponds to a base state $\ket{e}$ of the walker's Hilbert space $\sH = \mathrm{span}\{\ket{e} | e \in E\}$ and represents the walker at position $v$ facing towards the vertex $w$. A general pure state of the walker reads
$$
\ket{\psi} = \sum_{e \in E} \psi_{e} \ket{e}.
$$
The edges originating in a common vertex $v$ defines the vertex subspace $\sH_v$. The complete Hilbert space of the quantum walk can then be decomposed as $\sH = \bigoplus_{v\in V} \sH_v$.Note that since the graph is not regular (except for the case (4)), the Hilbert space cannot be decomposed into a tensor product of a position space and a coin space.

As the initial point of the walk we consider the corner vertex $0$ (depicted by green in Figure~\ref{fig:ladders}). The initial state is localized at this vertex, i.e. $\ket{\psi(0)} \in \sH_{0}$, and it can be written in the form
\begin{equation}
\label{init}
\ket{\psi(0)} = x \ket{(0,2)} + y \ket{(0,1)} + z \ket{(0,0)}, \quad |x|^2 + |y|^2 + |z|^2 = 1.
\end{equation}
We consider the coined discrete time quantum walk evolution where each step consists of three subsequent operations. First, the unitary reflecting ("flip-flop") shift operation
\begin{align*}
R = \sum_{e \in E_a} \ket{\bar{e}}\bra{e} + \sum_{e\in E_l} \ket{e}\bra{e},
\end{align*}
swaps amplitudes of arcs going in opposite directions and leaves the amplitudes of loops unchanged. Next, the unitary coin operation $C = \bigoplus_{v\in V} C_v$ is performed which locally shuffles amplitudes in each individual vertex subspace $\sH_v$. As in \cite{mares2020}  we chose the extensively studied Grover coin 
$$
    C_v = 2\ket{\Psi_v}\bra{\Psi_v} - I_v,
$$
for all vertices. Here $I_v$ is the identity operator in $\sH_v$ and the state $\ket{\Psi_v}$ is the equal-weight superposition of all basis states of the standard basis of $\sH_v$, i.e.
$$
    \ket{\Psi_v} = \frac{1}{\sqrt{\mathrm{deg}(v)}}\sum_{\begin{matrix}
    v'\in V \\ (v,v')\in E \end{matrix}} \ket{(v,v')} .
$$
We have denoted by $\mathrm{deg}(v)$ the degree of the vertex $v\in V$. These two operations together form a unitary evolution  operator of the Grover walk
$$
U = C R.
$$
For consistency, we kept the order of the shift and the coin as in \cite{mares2020}. Note that quite often in the literature the reverse order is used. However, as argued in \cite{mares2019} the two variants are equivalent up to an additional application of $C^\dagger$ on the initial state and $C$ on the final state after $t$ steps.

Finally, at the end of each step, part of the walker's wave function which has reached the sink is eliminated. We consider the sink to be located at the corner vertex opposite the initial vertex across the ladder, i.e. the vertex $2L+1$ (depicted by red in Figure~\ref{fig:ladders}). The projector onto the sink subspace $\sH_{2L+1}$ is the operator $I_{2L+1}$. The walk continues only if the excitation is not absorbed in the sink, which we describe by the projection onto the complement of the sink subspace $\sH_{2L+1}$
$$
\Pi = I - I_{2L+1}.
$$ 
Altogether, single step of the coined quantum walk with sink is described by a non-unitary operator $\Pi U$. The non-normalized state of the walk after $t$ steps is given by
$$
\ket{\psi(t)} = (\Pi U)^t \ket{\psi(0)}.
$$
The square norm of this vector is the probability that the walker was not yet absorbed in the sink, i.e. the survival probability until time $t$
\begin{equation}
\label{surv:t}
S(t) = \lVert \psi(t) \rVert^2 = \braket{\psi(t)|\psi(t)}. \end{equation}
It can be shown in general that the limit of the survival probability exists \cite{KonnoSegawaStefanak}, let us denote the limit by $S$. In contrast to the classical random walk, the survival probability can have a non-vanishing limit due to the presence of the dark states. The dark states are eigenstates of the unitary operator $U$ which are not supported on the sink vertex. Hence, they are not affected by the sink projector $\Pi$, which means they are eigenstates of the non-unitary operator $\Pi U$ describing the evolution of the walk with an absorbing sink. The evolution in the dark subspace remains unitary. This is indeed the case of the ladder graph, as was identified in \cite{mares2020}. 

\begin{figure}[t]
    \centering
    \includegraphics[width=0.4\textwidth]{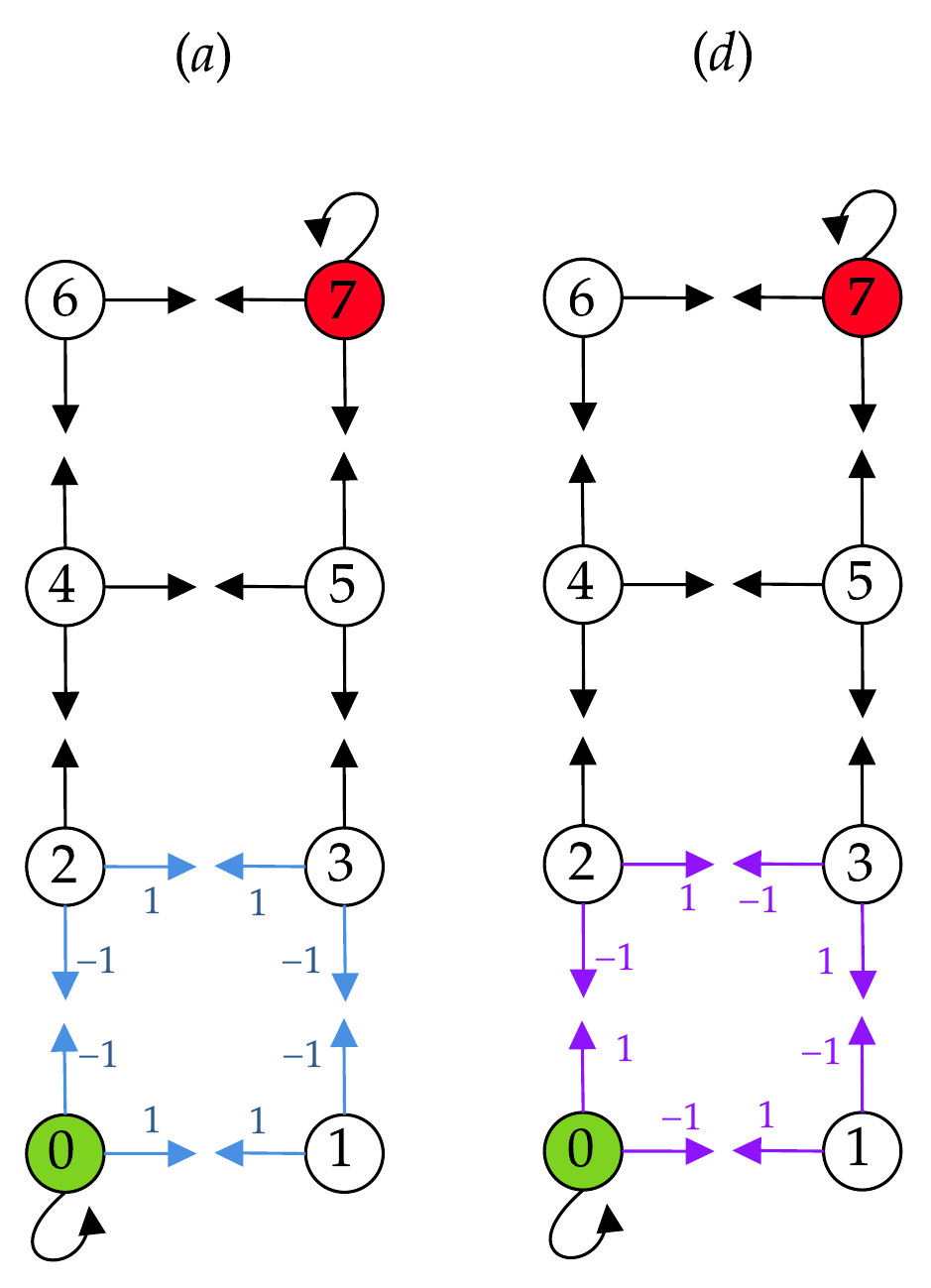}
    \caption{Linearly independent dark states of the Grover walk on the first configuration of the ladder graph. For $L=3$ there are $2$ face-cycle~states of types (a) and (d). Weights of the graphs are nonzero coefficients of the corresponding (non-normalized) dark states $\ket{a_i}$ and $\ket{d_i}$, given by equations (\ref{def:ai}) and (\ref{def:di}).}
    \label{fig:trapped_no_loops}
\end{figure}

We denote the dark subspace by ${\cal H}^D$. In general, for a Grover walk on a finite graph with sinks the dark subspace can be decomposed into a direct sum \cite{KonnoSegawaStefanak}
$$
\mathcal{H}^D = \mathcal{K}\oplus \mathcal{M}\oplus \mathcal{T},
$$
where $\mathcal{K}$ is a subspace of dark states corresponding to the eigenvalue 1, subspace $\mathcal{M}$ belongs to the eigenvalue -1 and the vectors from $\mathcal{T}$ have eigenvalues $\lambda\neq \pm 1$. As we prove in the \ref{app}, for the ladder graph the $\mathcal{T}$ subspace is empty. The vectors spanning $\mathcal{K}$ and $\mathcal{M}$ were identified in \cite{mares2020}. 
In the first configuration there are two types of face-cycle dark states, depicted in Figure~\ref{fig:trapped_no_loops}. The (a) type dark states correspond to the eigenvalue $-1$ of the operator $\Pi U$ (i.e., they belong to the $\mathcal{M}$ subspace) and have the form (for $i=0,\ldots, L-2$)
\begin{align}
\nonumber \ket{a_i} = \frac{1}{\sqrt{8}} \big( & - \ket{(2i,2i+2)} - \ket{(2i+2,2i)} + \ket{(2i+2,2i+3)} + \\
\nonumber & + \ket{(2i+3,2i+2)} - \ket{(2i+3,2i+1)} -  \ket{(2i+1,2i+3)} + \\
& + \ket{(2i+1,2i)} + \ket{(2i,2i+1)} \big) .
 \label{def:ai}
\end{align}
Note that for $i=L-1$ the formula yields an eigenvector of $U$ which, however, is supported at the sink vertex $2L+1$. Hence, this vector contributes to quantum transport and it is not a dark state. The $L-1$ vectors $\ket{a_i}, i=0,\ldots, L-2$ are linearly independent, however, consequent states are not orthogonal. In addition, there are $L-1$ linearly independent dark states of the type (d) (for $i=0,\ldots, L-2$)
\begin{align}
\nonumber \ket{d_i} = \frac{1}{\sqrt{8}} \big( & \ket{(2i,2i+2)} - \ket{(2i+2,2i)} + \ket{(2i+2,2i+3)} - \\ \nonumber & -\ket{(2i+3,2i+2)} + \ket{(2i+3,2i+1)} -  \ket{(2i+1,2i+3)} + \\
& + \ket{(2i+1,2i)} - \ket{(2i,2i+1)}\big ).
\label{def:di}
\end{align}
These eigenvectors correspond to the eigenvalue $1$ of $\Pi U$, i.e. vectors $\ket{d_i}$ belong to the $\mathcal{K}$ subspace and are orthogonal to $\ket{a_i}$ dark states. However, consequent dark states $\ket{d_i}$ are not mutually orthogonal.

\begin{figure}[t]
    \centering
    \includegraphics[width=0.4\textwidth]{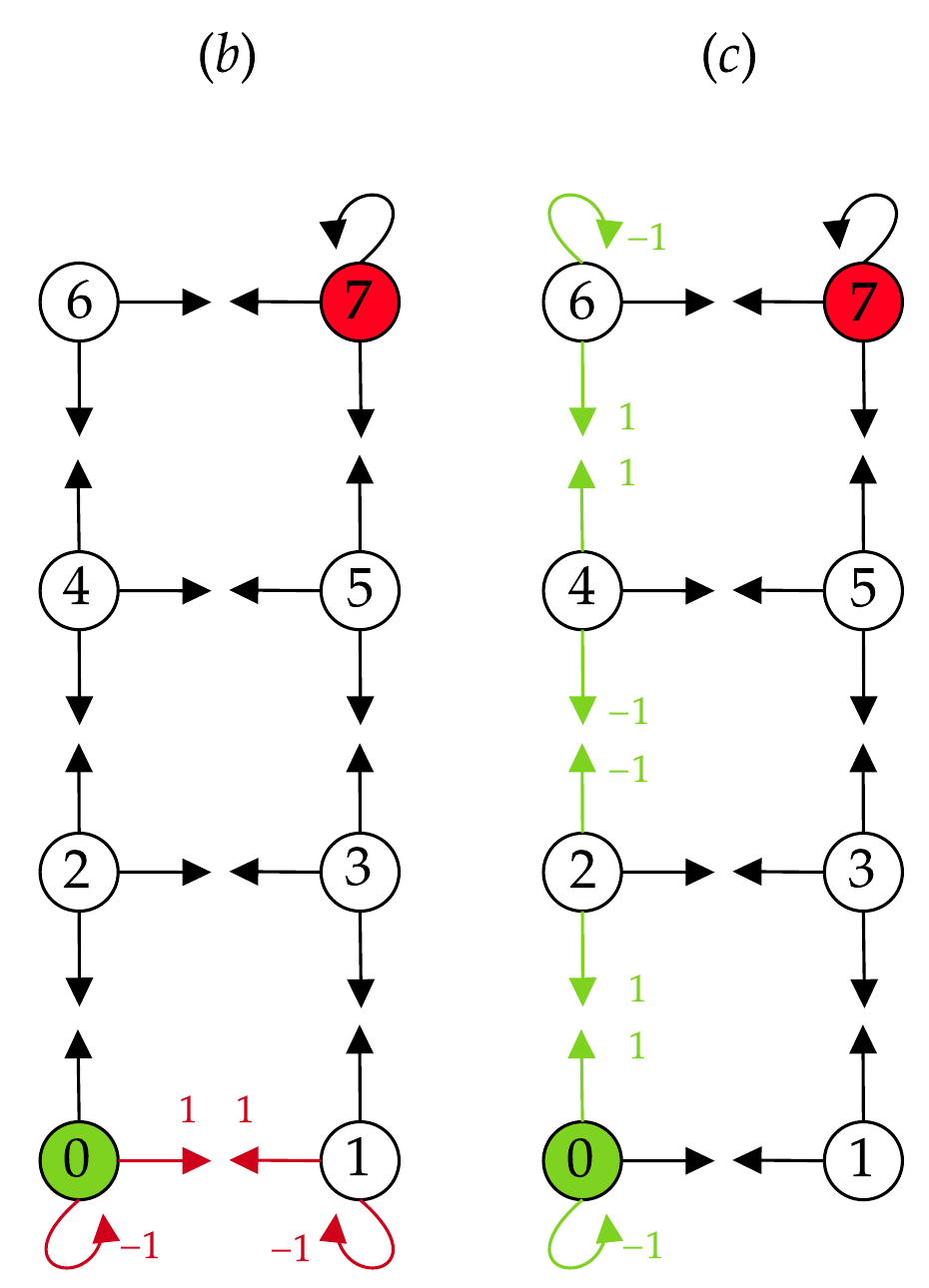}
    \caption{Additional linearly independent dark states of the Grover walk arising from extra loops. The loop at the bottom right corner adds the short-path dark state $\ket{b}$. The loop at the top left corner adds the long-path dark state $\ket{c}$. Weights of the graphs are nonzero coefficients of the corresponding (non-normalized) dark states, given by equations (\ref{def:b}) and (\ref{def:c}).}
    \label{fig:trapped_loops}
\end{figure}

When the loops are attached to the remaining corners of the ladder graph, two new dark states arise. We depict them in Figure~\ref{fig:trapped_loops}. With the addition of the loop to the bottom right corner, the short-path (or (b) type) dark state emerges
\begin{equation}
    |b\rangle = \frac{1}{2}\left(-|(0,0)\rangle + | (0, 1) \rangle + | (1, 0) \rangle -| (1,1) \rangle\right).
\label{def:b}
\end{equation}
Finally, the loop in the top left corner adds the long-path (or (c) type) dark state
\begin{align}
\nonumber |c\rangle =  \frac{1}{\sqrt{2(L+1)}} {\big [} &
- |(0,0) \rangle +(-1)^{L}|(2L,2L)\rangle + \\ 
& +\sum_{j=0}^{L-1}(-1)^{j}\left( \;|(2j,2j+2)\rangle + |(2j+2,2j)\rangle\;\right) {\big ]} .
\label{def:c}
\end{align}
Both $\ket{b}$ and $\ket{c}$ correspond to the eigenvalue $-1$ of the evolution operator $\Pi U$, i.e. they belong to the $\mathcal{M}$ subspace. They are orthogonal to the $\ket{d_i}$ dark states, but not with the $\ket{a_i}$ states, and neither they are mutually orthogonal.

We see that the $\cal K$ subspace is the same for all configurations of the ladder, since the $(d)$ type dark states are the only vectors corresponding to the eigenvalue 1. On the other hand, $\cal M$ subspaces differs. In the first configuration it is spanned solely by $(a)$ type states. In the second (third) configuration we have to add the $\ket{b}$ ($\ket{c}$) state, respectively. Finally, in the fourth configuration, we have to add both $\ket{b}$ and $\ket{c}$ to $\cal A$. For this reason, the survival probability depends on the configuration of the ladder graph. Indeed, the limiting value of the survival probability (\ref{surv:t}) is given by
\begin{equation}
S = \lVert \Pi_{\mathcal{K}}\psi(0) \rVert^2 + \lVert \Pi_{\mathcal{M}}\psi(0) \rVert^2 ,
\label{sp:general}
\end{equation}
where we have denoted the orthogonal projectors on the respective dark subspaces as $\Pi_\mathcal{K}$ and $\Pi_\mathcal{M}$. These orthogonal projectors are constructed in Sections~\ref{sec:3} and \ref{sec:4} by orthogonalization of the basis vectors (a)-(d) obtained from \cite{mares2020}. First, in Section~\ref{sec:3} we investigate the contribution of the face-cycle dark states $\ket{a_i}$ a $\ket{d_i}$. Second, in Section~\ref{sec:4} we focus on the contribution of the loops and resulting dark states $\ket{b}$ and $\ket{c}$.

%%%%%%%%%%%%%%%%%%%%%%%%%%%%%%%%%
%%%%%%%%%%%%%%%%%%%%%%%%%%%%%%%%%
\section{Contribution of face-cycles}
\label{sec:3}
%%%%%%%%%%%%%%%%%%%%%%%%%%%%%%%%%
%%%%%%%%%%%%%%%%%%%%%%%%%%%%%%%%%

In this section we consider the first version of the ladder graph. The dark eigenspaces are spanned solely by the face-cycles of the graph, i.e.
\begin{align}
    \nonumber \mathcal{K} & = \mathrm{span}\{ \ket{d_k}|k=0,\ldots, L-2 \} , \\
     \nonumber \mathcal{M} & \equiv \mathcal{A} =  \mathrm{span}\{ \ket{a_k}|k=0,\ldots, L-2 \} . 
\end{align}
Hence, in the first configuration of the ladder graph the  survival probability (\ref{sp:general}) is given by
$$
S_1(L) = S_d(L) + S_a(L),
$$
where we have denoted
    \begin{align*} 
    S_d(L) & = \lVert \Pi_{\mathcal{K}}\psi(0) \rVert^2 = \langle \psi(0)|\Pi_{\mathcal{K}}|\psi(0)\rangle , \\
    S_a(L) & = \lVert \Pi_{\mathcal{A}}\psi(0) \rVert^2 = \langle \psi(0)|\Pi_{\mathcal{A}}|\psi(0)\rangle
    \end{align*}
In the following we determine $S_a(L)$ and $S_d(L)$ as functions of the ladder length $L$ and the initial state $\ket{\psi(0)}$. To this end, we need to construct orthonormal bases in the subspaces ${\cal A}$ and $\cal K$. Let us begin with the subspace ${\cal A}$, the procedure is the same for $\cal K$. The linearly independent vectors (\ref{def:ai}) are not orthogonal, however, for the scalar product holds
\begin{equation}
\label{scalar:prod:a}
\langle a_i|a_j\rangle = \left\{\begin{array}{cc}
     \frac{1}{4},  & j = i\pm 1  \\ \\
    0,  & |i-j|>1
\end{array}\right.    
\end{equation}
We can employ this fact to construct an orthonormal set $\left\{\ket{\Phi_i},i=0,\ldots,L-2\right\}$ such that 
\begin{eqnarray}
\nonumber \ket{a_0} & = & \ket{\Phi_0} , \\
\label{a:phi} \ket{a_k} & = & \alpha_k \ket{\Phi_{k-1}} + \beta_k \ket{\Phi_{k}}, k=1,\ldots ,L-2 .
\end{eqnarray}
Since $\ket{a_i}$ are real valued vectors we can assume so are $\ket{\Phi_i}$ and therefore also the amplitudes $\alpha_k$ and $\beta_k$ are real. Hence, we have to find two sequences $(\alpha_k)$ and $(\beta_k)$ satisfying the relations 
\begin{eqnarray}
\beta_k \alpha_{k+1} & = & \frac{1}{4}, \label{eq:bkak}\\
\nonumber \beta_k^2 & = & 1-\alpha_k^2, 
\end{eqnarray}
which arise from the scalar product (\ref{scalar:prod:a}) and normalization, with the initial condition
\begin{equation}
\label{rec:init:a}
    \alpha_0 = 0,\quad \beta_0 = 1. 
\end{equation}
This leads us to the recurrence relation
\begin{equation}
\label{rec:a}
    \alpha_{k+1}^2 = \frac{1}{16(1-\alpha_k^2)}.
\end{equation}
Let us define 
$$
M = \begin{pmatrix}
\alpha & \beta \\
\gamma & \delta 
\end{pmatrix},
$$
and the map $\Gamma_M$ as
$$
\Gamma_M(z) = \frac{\alpha z + \beta}{\gamma z + \delta}. 
$$
The recurrence (\ref{rec:a}) corresponds to the matrix
$$
M = \begin{pmatrix}
0 & 1/16\\
-1 & 1 
\end{pmatrix}.
$$
Hence, we find
\begin{align*}
\alpha_{k}^2 &= \Gamma_M(\alpha_{k-1}^2) = \Gamma_{M^k}(\alpha_0^2) = \Gamma_{M^k}(0) = \frac{\left(M^k\right)_{1,2}}{\left(M^k\right)_{2,2}}  =  \frac{1}{4}\frac{U_{k-1}(2)}{U_k(2)}, 
\end{align*}
where $U_k(z)$ is the Chebyshev polynomial of the second kind, which satisfies the recurrence relation 
\begin{align}
    U_0(z)&=1,\;U_1(z)=2z \notag \\
    U_{n+1}(z) &= 2z U_n(z) - U_{n-1}(z). \label{rec:cheb}
\end{align}
We put $U_k:=U_k(2)$. 
An explicit expression of  $U_k$ is  
\begin{equation}
\label{def:Uk}
    U_k=\frac{\lambda_+^{k+1}-\lambda_-^{k+1}}{\lambda_+-\lambda_-}>0,
\end{equation}
where $\lambda_+>\lambda_-$ are the roots of the quadratic equation
$$
\lambda^2 - 4\lambda + 1 = 0, 
$$
i.e., $\lambda_{\pm}=2\pm \sqrt{3}$. Note that the roots satisfy conditions
\begin{equation}
\label{roots:cond}
    \lambda_+\lambda_- = 1 , \quad \lambda_+ +\lambda_- = 4 .
\end{equation}
After all, we find that the squares of the amplitudes are given by
\begin{align*}
   \alpha_k^2 & = \frac{1}{4}\frac{U_{k-1}}{U_k}, \quad \beta_k^2  = 1 - \frac{1}{4}\frac{U_{k-1}}{U_k} = \frac{1}{4}\frac{U_{k+1}}{U_k}, 
   \end{align*}
where we have used the recurrence (\ref{rec:cheb}) for $z=2$.
Since the global phase of the quantum state $\ket{\Phi_k}$ has no physical meaning, only the relative phase between the amplitudes $\alpha_k$ and $\beta_k$ is relevant. 
In accordance with the initial condition (\ref{rec:init:a}) we can choose $\beta_k>0$ for $k\geq 0$. It follows from (\ref{eq:bkak}) that $\alpha_{k+1}>0$ as well. Hence, the amplitudes are 
\begin{equation}
\label{ampl:abk}
       \alpha_k = \frac{1}{2}\sqrt{\frac{U_{k-1}}{U_k}}, \quad
   \beta_k = \frac{1}{2}\sqrt{\frac{U_{k+1}}{U_k}} .
\end{equation}

The projection operator onto the dark subspace $\mathcal{A}$ is given by
$$
\Pi_{\mathcal{A}} =\sum_{k=0}^{L-2}|\Phi_k \rangle \langle \Phi_k|. 
$$
However, to evaluate the application of the projector onto the initial state $|\psi(0) \rangle$ it is more convenient to express it in terms of the vectors $\ket{a_i}$. We derive this form in the following Lemma~\ref{lem:cycle}.

%%%%%%%%%%%%%%%%
\begin{lemma}
\label{lem:cycle}
Let $\Theta_{ij}$ be given by
\begin{equation}
    \Theta_{ij} = 4 (-1)^{i+j} U_i U_j \sum_{k=\max\{i,j\}}^{L-2} \frac{1}{U_k U_{k+1}} ,
    \label{theta:a}
\end{equation}
for $0\leq i,j\leq L-2$. The orthogonal projector on the subspace $\cal A$ is then given by
\begin{equation}
\label{proj:a}
  \Pi_{\mathcal{A}} = \sum_{i,j=0}^{L-2} \Theta_{ij} \ket{a_i}\bra{a_j}.  
\end{equation}
\end{lemma}
%%%%%%%%%%%%%%%%
\proof
Inverting the relations (\ref{a:phi}) we find the vectors of the orthonormal set $\ket{\Phi_k}$ in terms of the states $\ket{a_l}$
\begin{eqnarray}
\nonumber \ket{\Phi_k} & = & \frac{1}{\beta_k}\sum_{l=0}^k (-1)^{k-l} \left(\prod\limits_{i=l+1}^k \frac{\alpha_i}{\beta_{i-1}}\right) \ket{a_l} .
\end{eqnarray}
Using the explicit form of the amplitudes (\ref{ampl:abk}) we simplify the product
$$
\prod\limits_{i=l+1}^k \frac{\alpha_i}{\beta_{i-1}} = \prod\limits_{i=l+1}^k \frac{U_{i-1}}{U_i} = \frac{U_l}{U_k},
$$
which leads us to
\begin{equation}
\ket{\Phi_k} = \frac{2}{\sqrt{U_k U_{k+1}}}\sum_{l=0}^k (-1)^{k-l} U_l \ket{a_l} .
\label{phi:a}
\end{equation}
The projector onto the ${\cal A}$ subspace is then equal to
$$
\Pi_{\mathcal{A}} = \sum_{k=0}^{L-2} \frac{4}{U_k U_{k+1}}\sum_{l,m=0}^k  (-1)^{l+m} U_l U_m \ket{a_l}\bra{a_m},
$$
which is readily rewritten into the form (\ref{proj:a}). \qed

Construction of the orthonormal basis in the dark subspace $\mathcal{K}$ proceeds along the same lines. From (\ref{def:di}) we find that the scalar products of the dark states $\ket{d_i}$ read
$$
\langle d_i|d_j\rangle = \left\{\begin{array}{cc}
     -\frac{1}{4},  & j = i\pm 1  \\ \\
    0,  & |i-j|>1
\end{array}\right.    
$$
We denote the orthonormal set spanning $\mathcal{K}$ as $\left\{\ket{\Phi_i'},i=0,\ldots,L-2\right\}$. In analogy with (\ref{a:phi}) we have the relation between $\ket{d_i}$ and   $\ket{\Phi_i'}$
\begin{eqnarray}
\nonumber \ket{d_0} & = & \ket{\Phi_0'} , \\
\nonumber \ket{d_k} & = & \alpha_k' \ket{\Phi_{k-1}'} + \beta_k' \ket{\Phi_{k}'},\quad  k=1,\ldots ,L-2 .
\end{eqnarray}
The amplitudes $\alpha_k'$ and $\beta_k'$ satisfy relations
\begin{eqnarray}
\nonumber \beta_k' \alpha_{k+1}' & = & -\frac{1}{4}, \quad  \beta_k'^2  =  1-\alpha_k'^2,
\end{eqnarray}
with initial conditions
$$
\alpha_0' = 0,\quad \beta_0' = 1.
$$
Choosing $\beta_k'>0$ for $k\geq 0$ we find the solution
$$
\alpha_k' = - \frac{1}{2}\sqrt{\frac{U_{k-1}}{U_k}}, \quad
   \beta_k' = \frac{1}{2}\sqrt{\frac{U_{k+1}}{U_k}} .
$$
In the same way as in (\ref{phi:a}) we obtain the expression of the vectors from the othonormal set $\ket{\Phi_k'}$ in terms of $\ket{d_l}$
$$
\ket{\Phi_k'} = \frac{2}{\sqrt{U_k U_{k+1}}}\sum_{l=0}^k U_l \ket{d_l} .
$$
The projector onto the ${\cal K}$ subspace is then given by
\begin{equation}
\label{proj:d}
\Pi_{\mathcal{K}} = \sum_{i,j=0}^{L-2} \Theta_{ij}' \ket{d_i}\bra{d_j},
\end{equation}
where we have denoted
\begin{equation}
    \Theta_{ij}' = 4 U_i U_j \sum_{k=\max\{i,j\}}^{L-2} \frac{1}{U_k U_{k+1}} .
    \label{theta:d}
\end{equation}

We are now prepared to prove the following  proposition.
%%%%%%%%%%%%%%%%%%%

\begin{theorem}
\label{thm:sigmaad}
For the initial state of the form (\ref{init}) the contribution of the cycles to the survival probability is given by
$$
S_a(L) = S_d(L) = \frac{1}{2} |x-y|^2 \Sigma(L),
$$
where we have denoted 
$$
    \Sigma(L) = \sum\limits_{k=0}^{L-2}\frac{1}{U_k U_{k+1}}.
$$
\end{theorem}
%%%%%%%%%%%%%%%%%%%
\proof
The support of the initial state $\ket{\psi(0)}$ is included in that of $\ket{a_0}$ and $\ket{d_0}$. From the explicit form of the vectors $\ket{a_i}$ (\ref{def:ai}) and $\ket{d_i}$ (\ref{def:di}) we find that
\begin{equation}
 \langle a_i|\psi(0)\rangle = -\langle d_i|\psi(0)\rangle = \frac{1}{\sqrt{8}} (y-x) \delta_{i,0}.
\label{scp:init:a}   
\end{equation}
Together with (\ref{proj:a}) and (\ref{proj:d}) this implies 
\begin{eqnarray}
\nonumber S_a(L) & = & \langle \psi(0)|\Pi_{\mathcal{A}}|\psi(0)\rangle =  |\langle a_0|\psi(0)\rangle|^2 \Theta_{00} = \frac{1}{8}|x-y|^2 \Theta_{00}, \\
\nonumber S_d(L) & = & \langle \psi(0)|\Pi_{\mathcal{K}}|\psi(0)\rangle  =  |\langle d_0|\psi(0)\rangle|^2 \Theta_{00}' = \frac{1}{8}|x-y|^2 \Theta_{00}'.
\end{eqnarray}
Finally, from (\ref{theta:a}) and (\ref{theta:d}) we obtain
\begin{equation}
\Theta_{00} = 4 \Sigma(L) = \Theta_{00}'. 
\label{theta:sigma}
\end{equation}

\qed

Let us now estimate $\Sigma(L)$. 

%%%%%%%%%%%%%%%%%%%
\begin{lemma}\label{lem:sigmaL}
For large $L$ the function $\Sigma(L)$ converges exponentially fast to a constant, i.e.
$$
\Sigma(L) \asymp 1-\lambda_-^{2(L+1)} 
$$
Here $f(x) \asymp g(x)$ means $0<\lim\limits_{x\to\infty} f(x)/g(x)<\infty$.
\end{lemma}
%%%%%%%%%%%%%%%%%%%%
\proof
{\bf Upper bound of $\Sigma(L)$:}  
From the expression (\ref{def:Uk}) and the fact that $\lambda_+>1>\lambda_->0$, we find $U_{k+1}>U_k$. Hence
\begin{align*}
    \Sigma(L) 
    &\leq \sum_{k=0}^{L-2} \frac{1}{U_{k}^2} = \sum_{k=0}^{L-2} \frac{(\lambda_+-\lambda_-)^2}{(\lambda_+^{k+1}-\lambda_-^{k+1})^2} = \sum_{k=0}^{L-2} \frac{(\lambda_+-\lambda_-)^2}{\lambda_+^{2(k+1)}(1-\lambda_-^{2(k+1)})^2} \\ 
    &\leq \frac{(\lambda_+-\lambda_-)^2}{(1-\lambda_-^2)^2\lambda_+^2}\sum_{k=0}^L \frac{1}{\lambda_+^{2k}}  =  \frac{1-\lambda_-^{2(L+1)}}{1-\lambda_-^2} \\
    & < \frac{1}{1-\lambda_-^2}=1.07735\cdots ,
\end{align*}
where we have used the relations (\ref{roots:cond}) in the final equation. Then $\Sigma(L)$ is uniformly bounded. 
Since $\Sigma(L)$ is monotone increasing with respect to $L$, the limit $\lim\limits_{L\to\infty}\Sigma(L)$ exists. \\
{\bf Lower bound of $\Sigma(L)$:} 
We can compute that 
\begin{align*}
    \Sigma(L) 
    &\geq \sum_{k=0}^L \frac{1}{U_{k+1}^2} = \sum_{k=0}^L \frac{(\lambda_+-\lambda_-)^2}{\lambda_+^{2(k+2)}(1-\lambda_-^{2(k+2)})^2} \\
    & \geq \frac{(\lambda_+-\lambda_-)^2}{\lambda_+^4}\sum_{k=0}^L \frac{1}{\lambda_+^{2k}} = \frac{(\lambda_+-\lambda_-)^2}{\lambda_+^4} \frac{1-\lambda_-^{2(L+1)}}{1-\lambda_-^2} \\
    & > \lambda_-^2-\lambda_-^4 = 0.066642\cdots .
\end{align*}
Then $\lim\limits_{L\to\infty} \Sigma(L)$ is bounded as follows. 
    $$ 
    \lambda_-^2-\lambda_-^4< \lim_{L\to\infty}\Sigma(L) < \frac{1}{1-\lambda_-^2} 
    $$
Numerical evaluation reveals that the limiting value can be estimated by
$$
\lim_{L\to\infty}\Sigma(L) = \Sigma(\infty) \approx 0.26795 .
$$
\qed

Combining Theorem~\ref{thm:sigmaad} with Lemma~\ref{lem:sigmaL}, we obtain the following estimation of $S_1(L)$.  
\begin{corollary}\label{cor:sigmaa}
The survival probability in the first configuration of the ladder graph equals
\begin{align*}
S_1(L) = S_a(L)+S_d(L)=|x-y|^2 \Sigma(L), 
\end{align*}
which is monotone increasing with $L$. For large $L$ it behaves as
\begin{equation}
S_1(L) \sim |x-y|^2 \Sigma(\infty) + O(\lambda_-^{2L}),
    \label{s1}
\end{equation}
where $a_n\sim b_n$ means $\lim_{n\to\infty}a_n/b_n=1$.
\end{corollary}

Without the loops at the bottom right and upper left corners the survival probability behaves as can be intuitively expected. Lengthening of the ladder increases the number of dark states, which in turn increases the survival probability. Nevertheless, with increasing length of the ladder the contribution of additional face-cycle dark states diminishes rapidly, and the survival probability saturates exponentially quickly with increasing $L$. However, if we add loops to the corners of the ladder graph this intuition fails. As we will see, the survival probability can become asymptotically decreasing; i.e. the ``Hashigo-sake" phenomenon of the quantum walk. In the next section, we mathematically clarify its emergence.  

\section{Contribution of loops}
\label{sec:4}

Let us now investigate the contribution of loops to the survival probability. Presence of loops results in the short and long-path dark states, which belong to the $\cal M$ subspace. To construct the orthogonal projector $\Pi_{\cal M}$ we have to add either $\ket{b}$ or $\ket{c}$ (or both, depending on the configuration of the ladder) to $\cal A$ with the Gram-Schmidt procedure.

%%%%%
\subsection{The pair of loops with short distance}

For the second configuration of the ladder we add $\ket{b}$ to $\cal A$ and decompose the $\cal M$ subspace
$$
{\cal M} = {\cal A}\oplus {\cal B}, \qquad {\cal B} = \mathrm{span}\{\ket{B}\}, \qquad \ket{B} = \frac{(1-\Pi_{\cal{A}})\ket{b}}{||(1-\Pi_{\cal{A}})\ket{b}||}
$$
In this way we obtain the survival probability in the form
$$
    S_2(L) = S_1(L) + S_{b\setminus a}(L),
$$
where we have denoted 
$$
 S_{b\setminus a}(L) =  \lVert \Pi_{\mathcal{B}}\psi(0) \rVert^2 = \left|\langle B|\psi(0)\rangle \right|^2 . 
$$
We determine this expression in the following theorem. 
\begin{theorem}
The contribution of the short-path dark state to the survival probability, under the existence of face-cycle states, is equal to
\begin{equation}
S_{b\setminus a}(L)= \frac{|x \Sigma(L) + y(1-\Sigma(L)) - z|^2}{4-2\Sigma(L)}  . 
\label{surv:b}
\end{equation}
For large $L$ it behaves like
\begin{equation}
S_{b\setminus a}(L)\sim  
   \frac{|x \Sigma(\infty) + y(1-\Sigma(\infty)) - z|^2}{4-2\Sigma(\infty)} + O(\lambda_-^{2L}). 
\label{surv:b:asymp}
\end{equation}
\end{theorem}
\proof
Let us first determine the explicit form of the vector $\ket{B}$. Using the projector $\Pi_{\cal A}$ (\ref{proj:a}) derived in Lemma~\ref{lem:cycle} we find
$$
(I - \Pi_{\mathcal{A}})\ket{b} = \ket{b} - \sum_{i,j=0}^{L-2} \Theta_{ij} \langle a_j|b\rangle \ket{a_i}  .
$$
The scalar product between $\ket{a_j}$ and $\ket{b}$ is readily obtained from the definitions (\ref{def:ai}) and (\ref{def:b})
$$
\langle a_j|b\rangle = \frac{1}{\sqrt{8}} \delta_{j,0},
$$
which leads us to
$$
(I - \Pi_{\mathcal{A}})\ket{b} = \ket{b} - \frac{1}{\sqrt{8}} \sum_{i=0}^{L-2} \Theta_{i0} \ket{a_i}  .
$$
The square norm of this vector is given by
$$
\lVert(I - \Pi_{\mathcal{A}})\ket{b} \rVert^2 = 1 - \langle b|\Pi_{\mathcal{A}}| b\rangle  = 1 - \frac{1}{8} \Theta_{00} = 1 - \frac{1}{2} \Sigma(L),
$$
where we have used (\ref{theta:sigma}). Hence, the vector $\ket{B}$ reads
\begin{equation}
\label{def:bp}
\ket{B} = \frac{1}{\sqrt{1-\frac{1}{2} \Sigma(L) }} \left(\ket{b} - \frac{1}{\sqrt{8}}\sum_{i=0}^{L-2} \Theta_{i0} \ket{a_i}  \right) .
\end{equation}
We determine the scalar product with the initial state (\ref{init})
$$
\braket{B|\psi(0)} = \frac{1}{\sqrt{1-\frac{1}{2} \Sigma(L) }} \left(\langle b|\psi(0)\rangle - \sqrt{2} \Sigma(L)\langle a_0|\psi(0)\rangle  \right) .
$$
Finally, using (\ref{scp:init:a}) and the fact that
$$
\langle b|\psi(0)\rangle = \frac{1}{2}(y-z) 
$$
we obtain
\begin{equation}
  \braket{B|\psi(0)} = \frac{1}{\sqrt{4-2\Sigma(L) }} \left(x \Sigma(L) + y(1-\Sigma(L)) - z\right),  
  \label{ov:B:psi0}
\end{equation}
which leads us to the formula (\ref{surv:b}) for the survival probability due to the $\ket{b}$ state. Since $\Sigma(L)$ has a limit $\Sigma(\infty)$, and the speed of convergence is estimated to be $O(\lambda_-^{2L})$ by Lemma~\ref{lem:sigmaL}, we immediately obtain the estimate (\ref{surv:b:asymp}) for large $L$.

\qed

Note that $S_{b\setminus a}(L)$ can decrease with $L$, e.g. if the initial state is $\ket{\psi(0)} = \ket{(0,1)}$ corresponding to $y=1$ and $x=z=0$. However, the decrease in $S_{b\setminus a}(L)$ is always smaller than the increase in $S_{1}(L)$. Hence, the total survival probability for the second configuration of the ladder $S_{2}(L)$ is increasing for all initial states. Moreover, the convergence speed remains exponential in $L$. We summarize this result in the following corollary.

\begin{corollary}
\label{cor:s2}
The survival probability for the second configuration of the ladder is monotone increasing and for large $L$ behaves as  
\begin{equation}
S_{2}(L)\sim |x-y|^2 \Sigma(\infty) +  
  \frac{|x \Sigma(\infty) + y(1-\Sigma(\infty)) - z|^2}{4-2\Sigma(\infty)} + O(\lambda_-^{2L}). 
  \label{s2}
\end{equation}
\end{corollary}

\begin{figure}
    \centering
    \includegraphics[width=0.5\textwidth]{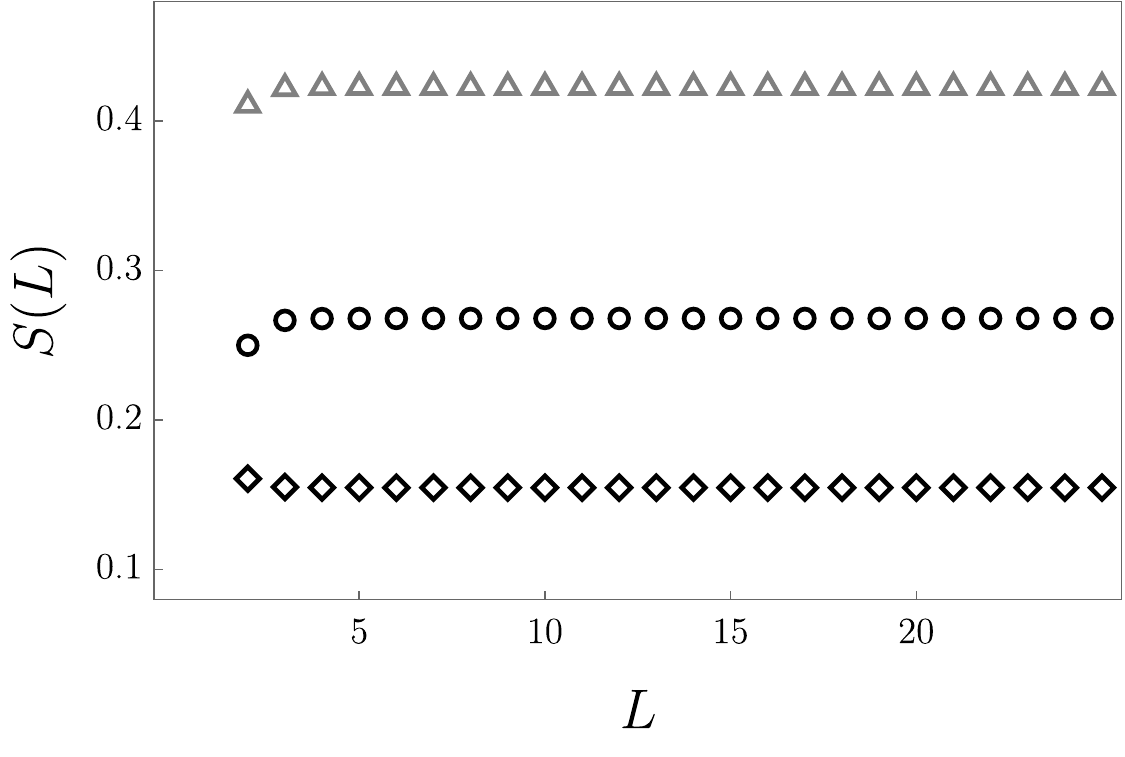}
    \caption{Comparison of the survival probability for the first and the second configuration of the ladder graph. $S_1(L)$ corresponds to black circles, $S_{b\setminus a}(L)$ to black diamonds and $S_2(L)$ is depicted by gray triangles. The initial state was chosen as $y=1$, $x=z=0$, i.e. $\ket{\psi(0)} = \ket{(0,1)}$. Both $S_1(L)$ and $S_{b\setminus a}(L)$ converge exponentially quickly to their limiting values (\ref{s1}) and (\ref{surv:b:asymp}). For the chosen initial state $S_{b\setminus a}(L)$ (\ref{surv:b}) is decreasing with $L$, however, the decrease is smaller than the increase in $S_1(L)$. Hence, $S_2(L)$ remains an increasing function of $L$.}
    \label{fig:s1s2}
\end{figure}

For illustration, we display in Figure~\ref{fig:s1s2} the survival probability for the first and the second configuration of the ladder when the initial state is given by $\ket{\psi(0)} = \ket{(0,1)}$.

%%%%%%%%%%%%%%%%%%%%%%%%%%%%%%%%%%%%%%%%%%%%%%%%%%%%%%

\subsection{The pair of loops with long distance}

For the third configuration of the ladder we have to add $\ket{c}$ to $\cal A$ and decompose the $\cal M$ subspace as
$$
{\cal M} = {\cal A}\oplus {\cal C}, \qquad {\cal C} =\mathrm{span}\{\ket{C}\}, \qquad \ket{C} = \frac{(1-\Pi_{\cal{A}})\ket{c}}{||(1-\Pi_{\cal{A}})\ket{c}||}.
$$
The corresponding survival probability is given by
$$
    S_3(L) =  S_1(L) + S_{c\setminus a}(L),
$$
where we have used the notation
$$
 S_{c\setminus a}(L) =  \lVert \Pi_{\mathcal{C}}\psi(0) \rVert^2 = \left|\langle C|\psi(0)\rangle \right|^2 . 
$$
We prove the following theorem:

\begin{theorem}
\label{thm:long}
The contribution of the long-path dark state to the survival probability, under the existence of face-cycle states, is equal to
\begin{equation}
\label{surv:c}
S_{c\setminus a}(L)=\frac{\left|x-z + \frac{y-x}{2}(1-\Sigma(L)-U_{L-1}^{-1})\right|^2}{\left( L+4-\Sigma(L)-U_{L-1}^{-1} \right)} .   
\end{equation}
For large $L$ we find the behaviour 
\begin{equation}
\label{surv:c:asymp}
S_{c\setminus a}(L)\sim \frac{\left|x(1+\Sigma(\infty)) + y(1-\Sigma(\infty)) - 2z\right|^2}{4\left( L+4-\Sigma(\infty)\right)} + O(\lambda_-^{2L}).
\end{equation}

\end{theorem}

\proof
It holds that 
    \begin{equation}
    \label{eq:sigmac}
    S_{c\setminus a}(L)=\frac{|\bra{c}(1-\Pi_{\cal{A}})\ket{\psi(0)}|^2}{||(1-\Pi_{\cal{A}})\ket{c}||^2}.
    \end{equation}
Now let us compute the numerator and the denominator, separately. First, using the projector (\ref{proj:a}) we obtain
$$
(1-\Pi_{\cal{A}})\ket{c} = \ket{c} - \frac{1}{2\sqrt{L+1}}\sum_{i,j=0}^{L-2} (-1)^{j+1} \Theta_{ij} \ket{a_i} 
$$
where we have used the scalar product between $\ket{a_j}$ and $\ket{c}$
\begin{equation}
\langle a_j|c\rangle = \frac{(-1)^{j+1}}{2\sqrt{L+1}} .
\label{ov:aj:c}
\end{equation}
The square norm of this vector reads
\begin{equation}
 \label{sum2}   ||(1-\Pi_{\cal{A}})\ket{c}||^2  = 1 - \bra{c}\Pi_{\cal A}\ket{c}  = 1 - \frac{1}{4(L+1)}\sum_{i,j=0}^{L-2} (-1)^{i+j+2} \Theta_{ij} .
\end{equation}
For the scalar product with the initial state we find
\begin{align}
 \nonumber    \bra{c}1-\Pi_{\cal{A}}\ket{\psi(0)} & = \langle c|\psi(0)\rangle - \frac{1}{2\sqrt{L+1}}\sum_{i,j=0}^{L-2} (-1)^{j+1} \Theta_{ij} \langle a_i|\psi(0)\rangle \\
 \label{sum1}   & = \frac{x-z}{\sqrt{2(L+1)}} - \frac{y-x}{4\sqrt{2(L+1)}}\sum_{j=0}^{L-2} (-1)^{j+1} \Theta_{0j},
\end{align}
where we have used (\ref{scp:init:a}). To proceed, we evaluate the sums in (\ref{sum2}) and (\ref{sum1}) in the following lemma.

\begin{lemma}
\label{lem:thetaL}
Let $\Theta_{ij}$ be defined as in (\ref{theta:a}). Then we have 
\begin{align*}
    \sum_{j=0}^{L-2} (-1)^{j+1} \Theta_{0j} &= -2(1-\Sigma(L)-U_{L-1}^{-1}) \\
    \\
     \sum_{i,j=0}^{L-2} (-1)^{i+j+2} \Theta_{ij} &= 2(L-2+\Sigma(L)+U_{L-1}^{-1}) . 
\end{align*}
\end{lemma}
\proof
Concerning the first equality, from Lemma~\ref{lem:cycle} we find that  
\begin{align*}
\sum_{j=0}^{L-2} (-1)^{j+1} \Theta_{0j}  &= -4\sum_{j=0}^{L-2} U_j \sum_{k=j}^{L-2} \frac{1}{U_k U_{k+1}} = -4\sum_{k=0}^{L-2} \frac{1}{U_k U_{k+1}} \sum_{j=0}^{k} U_{j} . 
\end{align*}
The recurrence relation for the Chebyshev polynomial (\ref{rec:cheb}) for $z=2$ implies
\begin{equation}
\sum_{j=0}^{k} U_{j} = \frac{1}{2}(U_{k+1} - U_k - 1) , 
\label{sum3}
\end{equation}
which leads us to the desired expression
\begin{align*}
\sum_{j=0}^{L-2} (-1)^{j+1} \Theta_{0j}  &= -2\sum_{k=0}^{L-2}\left(\frac{1}{U_k} - \frac{1}{U_{k+1}} - \frac{1}{U_k U_{k+1}} \right)  = -2(1-\Sigma(L)-U_{L-1}^{-1}). 
\end{align*}

For the proof of the second equality, using Lemma~\ref{lem:cycle} again, we have  
\begin{align*}
        \sum_{i,j=0}^{L-2} (-1)^{i+j+2} \Theta_{ij}
        & = 4\sum_{i,j=0}^{L-2} U_i U_j \sum_{k=\max\{i,j\}}^{L-2} \frac{1}{U_k U_{k+1}} = 4\sum_{k=0}^{L-2}\frac{1}{U_k U_{k+1}} \left(\sum_{i=0}^{k} U_i \right)^2 .\\
\end{align*}
Using (\ref{sum3}) we find
\begin{align*}
 \sum_{i,j=0}^{L-2} (-1)^{i+j+2} \Theta_{ij} & = 
     \sum_{k=0}^{L-2}\frac{(U_{k+1}-U_{k}-1)^2}{U_k U_{k+1}} \\
     & = \sum_{k=0}^{L-2}\left(\frac{U_{k+1}^2 + U_{k}^2 + 1}{U_k U_{k+1}} - \frac{2}{U_k} + \frac{2}{U_{k+1}} - 2 \right)\\
        & = 2\left(K(L)+\Sigma(L)-L+U_{L-1}^{-1}\right),
\end{align*}
where we have denoted 
$$
K(L)=\frac{1}{2}\sum_{k=0}^{L-2}\frac{U_{k+1}^2+U_{k}^2-1}{U_k U_{k+1}}.
$$
By the formula of the Chebyshev polynomial 
$$
U^2_k(z)-U_{k+1}(z)U_{k-1}(z)=1,
$$
and the recursion (\ref{rec:cheb}), we have 
    \begin{align*} 
    U_{k+1}^2(z) +U_k^2(z)-1  &=U_{k+1}(z)(2zU_k(z)-U_{k-1}(z))+U_k^2(z)-1 \\
    &= 2zU_{k+1}(z)U_k(z). 
    \end{align*}
Inserting $z=2$ into the above, we notice that $K(L)$ is simply reduced to 
$$
K(L)=2(L-1).
$$
Then we obtain the final expression
$$
    \sum_{i,j=0}^{L-2} (-1)^{i+j+2} \Theta_{ij}=2(L-2+\Sigma(L)+U_{L-1}^{-1}). 
$$   \qed

Returning back to the evaluation of the survival probability due to the $\ket{c}$ state (\ref{eq:sigmac}), the denominator (\ref{sum2}) can be expressed using Lemma~\ref{lem:thetaL} as follows
\begin{equation}
||(1-\Pi_{\cal{A}})\ket{c}||^2 = \frac{1}{2(L+1)}\left( L+4-\Sigma(L)-U_{L-1}^{-1} \right) .
\label{eq:denominatorc}
\end{equation}
The numerator is determined by the absolute value square of the scalar product (\ref{sum1}), which is rewritten using Lemma~\ref{lem:thetaL} into the form
\begin{equation}
\bra{c}1-\Pi_{\cal{A}}\ket{\psi(0)}  = \frac{1}{\sqrt{2(L+1)}}\left(x-z + \frac{y-x}{2}(1-\Sigma(L)-U_{L-1}^{-1})\right) .
\label{eq:numeratorc}
\end{equation}
Overall, by (\ref{eq:denominatorc}) and (\ref{eq:numeratorc}), we reach the expression (\ref{surv:c}). The behaviour (\ref{surv:c:asymp}) for large $L$ follows immediately from $U_L^{-1}=O(\lambda_-^{L})$ and Lemma~\ref{lem:sigmaL}.\qed

The result (\ref{surv:c}) shows that, unlike $S_1(L)$ and $S_{b\setminus a}(L)$, $S_{c\setminus a}(L)$ is always monotone decreasing with $L$. As discussed in \cite{mares2020}, the reason for this behaviour is the vanishing normalization factor of the long-path dark state (\ref{def:c}), which decreases with $L^{-\frac{1}{2}}$. Consequently, the contribution of $\ket{c}$ to the survival probability vanishes like $L^{-1}$, resulting in the ``Hashigo-sake" phenomenon of the quantum walk. Note, however, that there are exceptional initial states, for which $S_{c\setminus a}(L)$ decreases exponentially. In such a case, the survival probability remains monotone increasing, as we point out in the following corollary.

\begin{corollary}
\label{cor:s3}
The survival probability for the third configuration of the ladder is asymptotically decreasing and for large $L$ behaves as  
\begin{equation}
S_{3}(L)\sim |x-y|^2 \Sigma(\infty)  + \frac{\left|x(1+\Sigma(\infty)) + y(1-\Sigma(\infty)) - 2z\right|^2}{4\left( L+4-\Sigma(\infty)\right)} + O(\lambda_-^{2L}),
\label{s3}
\end{equation}
unless the amplitudes of the initial state satisfy the condition
$$
z = \frac{1}{2}\left(x(1+\Sigma(\infty)) + y(1-\Sigma(\infty))\right) .
$$
In the latter case $S_{3}(L)$ is monotone increasing and reduces to $S_{1}(L)$ for large $L$.
\end{corollary}

\begin{figure}
    \centering
    \includegraphics[width=0.48\textwidth]{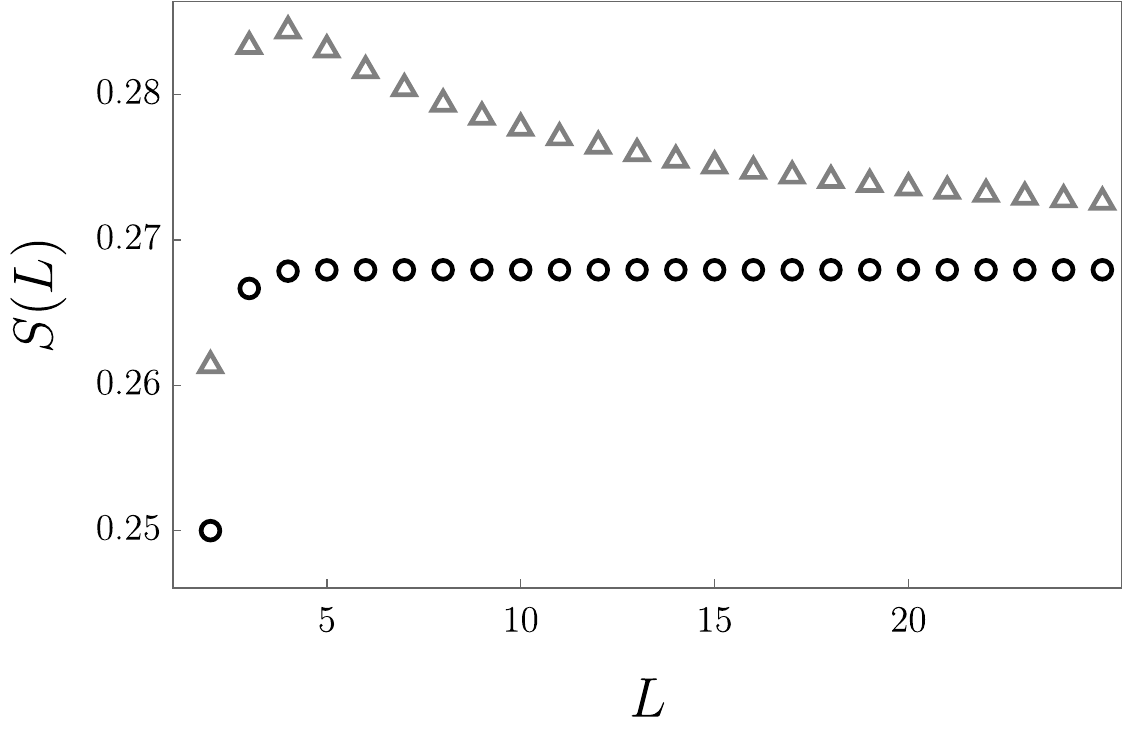}\hfill
    \includegraphics[width=0.48\textwidth]{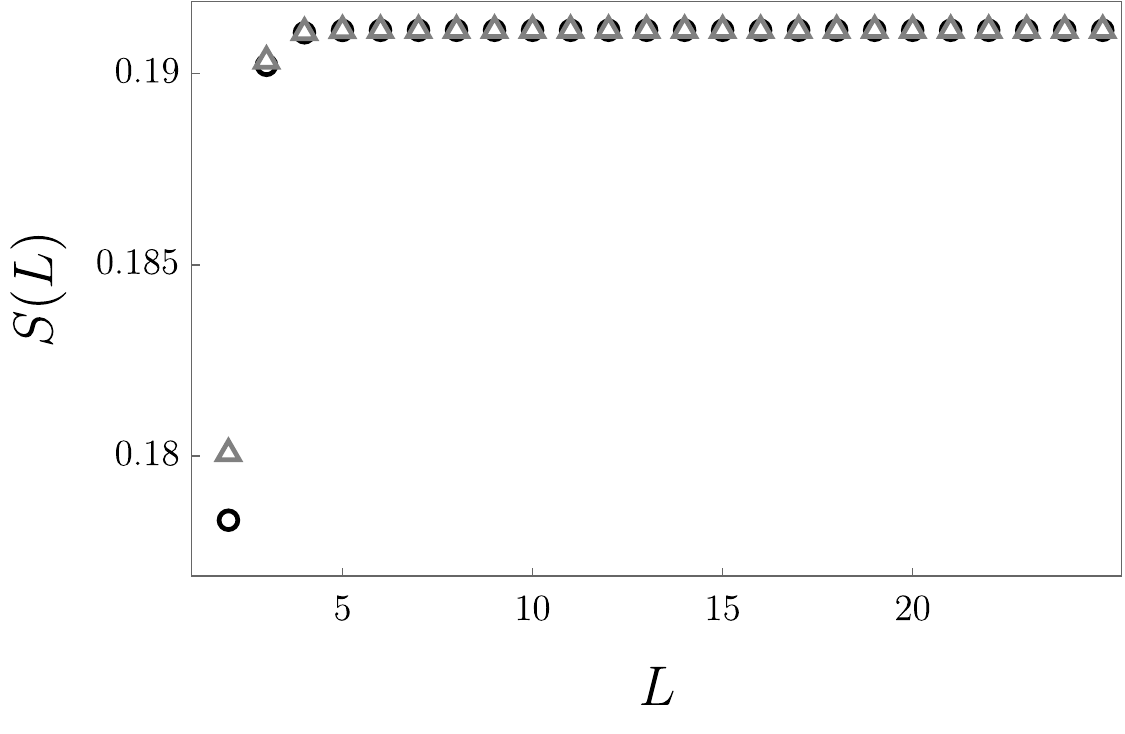}
    \caption{Comparison of the survival probability as a function of the ladder length $L$ for the first and the third configuration of the graph. $S_1(L)$ is depicted with black circles, $S_3(L)$ with gray triangles. In the left plot we have chosen the amplitudes of the initial state as $y=1$, $x = z =0$, corresponding to $\ket{\psi(0)} = \ket{(0,1)}$. $S_1(L)$ is increasing and exponentially quickly approaches the limiting value (\ref{s1}). For $L>4$ $S_3(L)$ decreases and the convergence speed is proportional to $L^{-1}$ according to (\ref{s3}). For the plot on the right the amplitudes of the initial state are given by $y=0$, $z = \frac{1}{2}x(1+\Sigma(\infty)) $, and $x = 2\left(5+2\Sigma(\infty) + \Sigma(\infty)^2\right)^{-1/2}$ to ensure proper normalization. In this case both $S_1(L)$ and $S_3(L)$ are increasing and converge quickly to (\ref{s1}), since the contribution of the long path dark states to the survival probability (\ref{surv:c:asymp}) vanishes exponentially.} 
    \label{fig:s1s3}
\end{figure}

We illustrate the role of the initial state on the survival probability in Figure~\ref{fig:s1s3}.

%%%%%%%%%%%%%%%%%%%%%%%%%%%%%%%%%%%%%%%%%%

\subsection{Two pairs of loops}

Finally, we consider the last configuration of the ladder. Performing the Gram-Schmidt procedure first for $\ket{b}$\footnote{As an alternative approach, we can choose the starting point from $\ket{c}$} and then for $\ket{c}$, we find the decomposition of dark subspace $\cal M$
$$
{\cal M} = {\cal A}\oplus {\cal B}\oplus {\cal C'}, \qquad {\cal C'} =\mathrm{span}\{\ket{C'}\}, \qquad \ket{C'} = \frac{(1-\Pi_{\cal{A}} - \Pi_{\mathcal B})\ket{c}}{||(1-\Pi_{\cal{A}} - \Pi_{\mathcal B})\ket{c}||} .
$$
The survival probability is then given by
$$
    S_4(L)  =  S_2(L) + S_{c\setminus\{a,b\}}(L),
$$
where we have denoted
$$
    S_{c\setminus\{a,b\}}(L) = \lVert \Pi_{\mathcal{C'}}\psi(0) \rVert^2 = \left|\langle C'|\psi(0)\rangle \right|^2 .
$$
We obtain the following expression for $S_{c\setminus\{a,b\}}(L)$. 

\begin{theorem}
\label{lem:sigmac-a,b}
The contribution of the long-path dark state to the survival probability, under the existence of face-cycle states and the short-path state, is equal to
\begin{equation}
\label{scab:full}
S_{c\setminus\{a,b\}}(L) = \frac{\left|x-z + \frac{2x-y-z}{(2-\Sigma(L))U_{L-1}} \right|^2}{4\left(L+4-\Sigma(L)-U_{L-1}^{-1} - \frac{(2 - \Sigma(L) - U_{L-1}^{-1})^2}{4-2 \Sigma(L)} \right)}. 
\end{equation}
For large $L$ it behaves like 
\begin{equation}
\label{scab:asymp}
S_{c\setminus\{a,b\}}(L)  \sim \frac{|x-z|^2}{4(L+3-\Sigma(\infty)/2)}+O(\lambda_-^{2L}) .
\end{equation}
\end{theorem}
\proof
We express $S_{c\setminus (a,b)}(L)$ in the form
\begin{align*}
    S_{c\setminus\{a,b\}}(L) 
    &= \frac{|\bra{c} (1-\Pi_{\cal{A}}-\Pi_{\cal{B}})\ket{\psi(0)}|^2}{||(1-\Pi_{\cal{A}}-\Pi_{\cal{B}})\ket{c}||^2} .
\end{align*}
The numerator is expanded as 
\begin{align}
\label{num:c}    
\bra{c} (1-\Pi_{\cal{A}}-\Pi_{\cal{B}})\ket{\psi(0)} &= \braket{c|1-\Pi_{\cal{A}}|\psi(0)}-\langle c|B\rangle \langle B|\psi(0)\rangle . 
\end{align}
From the explicit form of the vector $\ket{B}$ given by (\ref{def:bp}) we evaluate the scalar product
\begin{align}
    \nonumber \braket{c|B} & = \frac{1}{\sqrt{1-\frac{1}{2} \Sigma(L) }} \left(\braket{c|b} - \frac{1}{\sqrt{8}}\sum_{j=0}^{L-2} \Theta_{j0} \braket{c|a_j}  \right) \\
    \nonumber & = \frac{1}{\sqrt{1-\frac{1}{2} \Sigma(L) }} \left(\frac{1}{2\sqrt{2(L+1)}} - \frac{1}{2\sqrt{8(L+1)}}\sum_{j=0}^{L-2} (-1)^{j+1}\Theta_{j0}   \right) \\
\label{ov:c:B}    & = \frac{2 - \Sigma(L) - U_{L-1}^{-1}}{\sqrt{2(L+1)(4-2 \Sigma(L)) }} ,
\end{align}
where we have used (\ref{ov:aj:c}), Lemma~\ref{lem:thetaL} and
$$
\braket{c|b} = \frac{1}{2\sqrt{2(L+1)}}.
$$
Utilizing (\ref{ov:B:psi0}) and  (\ref{eq:numeratorc}) we find that the numerator (\ref{num:c}) reads
\begin{align}
  \label{ov:C:psi0} 
  \bra{c} (1-\Pi_{\cal{A}}-\Pi_{\cal{B}})\ket{\psi(0)}  & =  \frac{1}{2\sqrt{2(L+1)}}\left(x-z + \frac{2x-y-z}{(2-\Sigma(L))U_{L-1}} \right) .
\end{align}
Turning to the denominator, using (\ref{eq:denominatorc}) and (\ref{ov:c:B}) we find that it is given by
\begin{align}
\nonumber   ||(1-\Pi_{\cal{A}}-\Pi_{\cal{B}})\ket{c}||^2 & = ||(1-\Pi_{\cal{A}})\ket{c}||^2 - |\braket{c|B}|^2 \\
\nonumber & = \frac{\frac{}{} L+4-\Sigma(L)-U_{L-1}^{-1} - \frac{(2 - \Sigma(L) - U_{L-1}^{-1})^2}{4-2 \Sigma(L)}}{2(L+1)}.
\end{align}
Combining this result with (\ref{ov:C:psi0}) we arrive at the formula (\ref{scab:full}). For large $L$, we can replace $\Sigma(L)$ with $\Sigma(\infty)$ due to Lemma~\ref{lem:sigmaL} and neglect the exponentially decaying term $U_{L-1}^{-1}$, which leads us to the approximation (\ref{scab:asymp}).
 \qed

As in the third configuration of the ladder graph, the contribution of the $\ket{c}$ dark state vanishes with increasing $L$. The decrease is proportional to $L^{-1}$, unless we choose a specific initial state. This fact can alter the behaviour of the survival probability, as we summarize in the final corollary.

%%%%%
\begin{corollary}
\label{thm:main}
The survival probability for the fourth configuration of the ladder is asymptotically decreasing and for large $L$ behaves as  
\begin{multline} 
S_4(L)
\sim 
|x-y|^2\Sigma(\infty)\\
+\frac{|x\Sigma(\infty)+(1-\Sigma(\infty))y-z|^2}{4(1-\Sigma(\infty)/2)}\\ +\frac{|x-z|^2}{4(L+3-\Sigma(\infty)/2)}+O(\lambda_-^L),
\label{s4}
\end{multline}
except when the amplitudes of the initial state satisfy $x=z$. In the latter case $S_4(L)$ is monotone increasing and reduces to $S_2(L)$ for large $L$.
\end{corollary}

\begin{figure}
    \centering
    \includegraphics[width=0.48\textwidth]{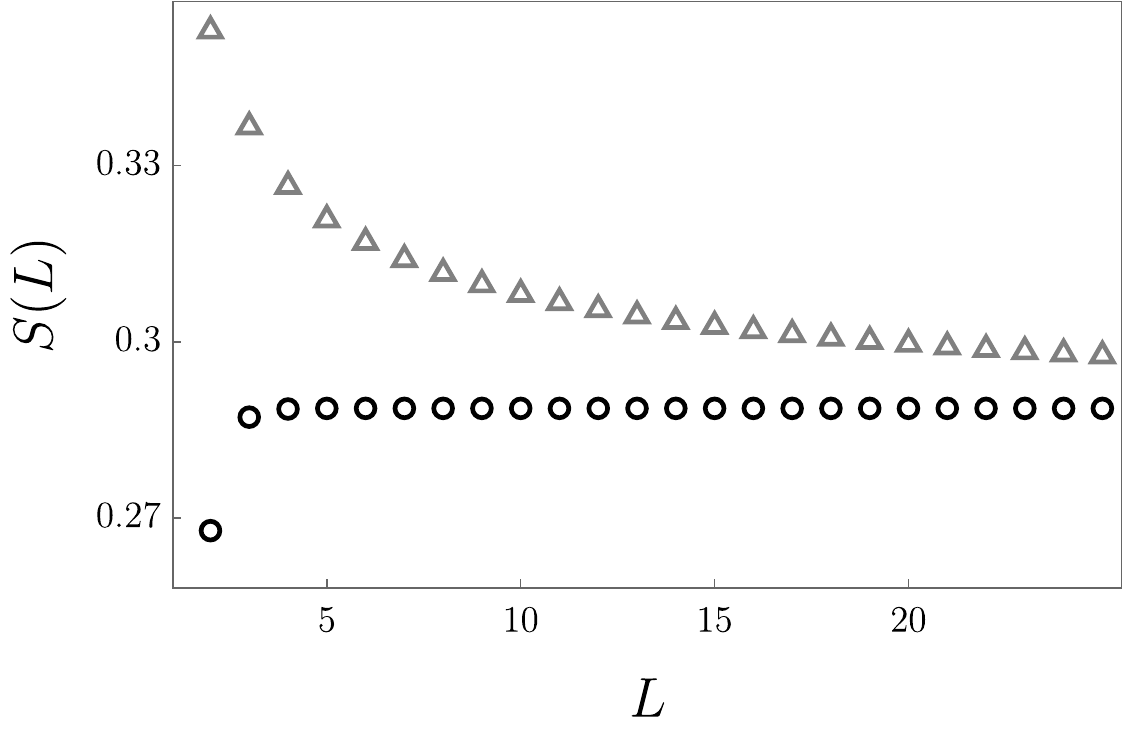}\hfill
    \includegraphics[width=0.48\textwidth]{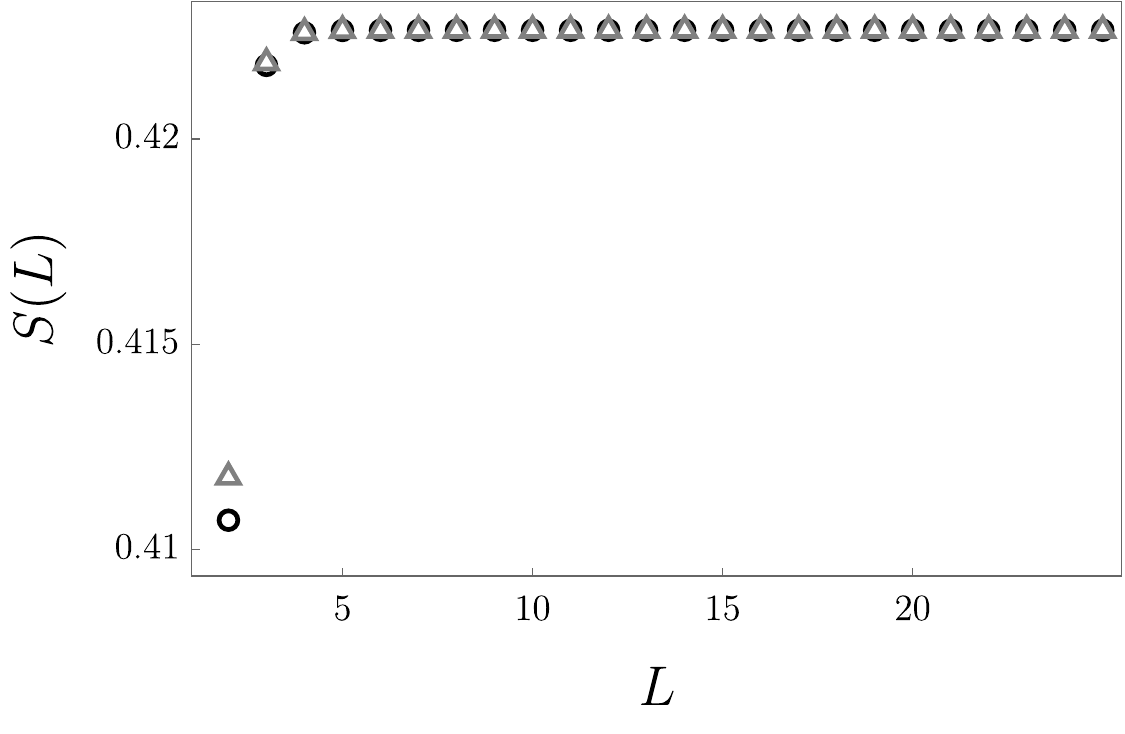}
    \caption{Comparison of the survival probability as a function of the ladder length $L$ for the second and the fourth configuration of the graph. $S_2(L)$ is depicted with black circles, $S_4(L)$ with gray triangles. In the left plot we have chosen the amplitudes of the initial state as $x=1$, $y = z =0$, corresponding to $\ket{\psi(0)} = \ket{(0,2)}$. $S_2(L)$ is increasing and exponentially quickly approaches the limiting value (\ref{s2}). $S_4(L)$ decreases with $L$ and the convergence speed is proportional to $L^{-1}$ according to (\ref{s4}). For the plot on the right the initial state is given by $y=1$, $x=z=0$, i.e. $\ket{\psi(0)} = \ket{(0,1)}$. In this case both $S_2(L)$ and $S_4(L)$ are increasing and converge quickly to (\ref{s2}), since the contribution of the long path dark states to the survival probability (\ref{scab:asymp}) vanishes exponentially. Note that for the third configuration of the ladder this initial state results in decreasing survival probability $S_3(L)$ (see the left plot in Figure~\ref{fig:s1s3}).} 
    \label{fig:s2s4}
\end{figure}

For illustration, we show the course of the survival probability as a function of $L$ for two different initial states in Figure~\ref{fig:s2s4}.

%%%%%%%%%%%%%%%%%%%%%%%%%%%%%%%%%%%%%%%%%%%%%%%%%%%%%%%%%%%%%%%%%%

\section{Conclusions}
\label{sec:concl}

We have investigated in detail the survival probability of the Grover walk on the ladder graph. The survival probability is obtained by the overlap between the initial state and the dark states. By taking the Gram-Schmidt procedure, we constructed orthonormal basis of the dark subspace and obtained a closed form of the survival probability. This allowed us to provide a rigorous explanation of the peculiar and counter-intuitive behaviour of the survival probability reported in \cite{mares2020}.

To explain the role of each graph structure for the survival probability we considered four configurations of the ladder, differing in the presence or absence of loops in the bottom right and upper left corner. In the first configuration without these loops the dark subspace is spanned solely by the face-cycles of the graph. We found that the survival probability is monotone increasing with the ladder size $L$. To some extent this matches a ``classical" intuition because the number of traps for the quantum walker represented by face-cycles in the graph increases with $L$. Adding a loop to the bottom right corner in the second configuration does not change the behaviour of the survival probability. Even though the contribution of the short-path dark state can be decreasing with $L$, the overall survival probability remains monotone increasing.

However, this changes dramatically when we consider the loop at the top left corner. The contribution of the long-path dark state is always decreasing with the ladder length, due to the vanishing normalization factor. Unless we choose a specific initial state, the decrease is inversely proportional to the ladder length $L$. This behaviour dominates and changes the survival probability to be asymptotically {\it decreasing} for the third and the fourth configuration of the ladder. Of course, just adding a loop cannot induce such a phenomenon in the corresponding random walk. Thus a counter intuition; ``Hashigo-sake" phenomenon of quantum walk, can be induced by only a small perturbation of the graph. Note that there is an interplay between the initial state and the graph configuration. As we have illustrated in Figures \ref{fig:s1s3} and \ref{fig:s2s4}, some initial states (e.g. $\ket{\psi(0)} = \ket{(0,1)}$) result in decreasing $S_3(L)$ but increasing $S_4(L)$ (or vice versa). The reason is indeed the non-orthogonality of the short-path and the long-path dark states.

Finally, let us compare the convergence speed of the survival probabilities. In the first two settings without the loop in the top left corner, a small size of the ladder, e.g. $L=5$, is enough for an estimation of the limiting value of the survival probability, since the convergence speed is exponential. However, in the third and the fourth setting the contribution of the long-path dark state to the survival probability is of the order of $O(L^{-1})$, except for special initial states. Hence, the convergence is {\it exponentially} slowed down by such a small perturbation of the graph, namely adding one loop at the top left corner. Thus, from view point of the quantum walk, this perturbation is no longer small. 

%\section*{Acknowledgements}
\ack
M\v S is grateful for financial support from M\v SMT RVO 14000 and the Czech Grant Agency project number GA\v CR 23-07169S. This publication was funded by the project ``Centre for Advanced Applied Sciences", Registry No.~CZ.$02.1.01/0.0/0.0/16\_019/0000778$, supported by the Operational Programme Research, Development and Education, co-financed by the European Structural and Investment Funds and the state budget of the Czech Republic. E.S. acknowledges financial supports from the Grant-in-Aid of
Scientific Research (C) Japan Society for the Promotion of Science (Grant No.~19K03616) and Research Origin for Dressed Photon.

\appendix

%%%%%%%%%%%%%%%%%%%%%%%%%%%%%%%
%%%%%%%%%%%%%%%%%%%%%%%%%%%%%%%
\section{No additional dark state}
\label{app}
%%%%%%%%%%%%%%%%%%%%%%%%%%%%%%
%%%%%%%%%%%%%%%%%%%%%%%%%%%%%%

%\normalsize

Let us prove that there are no additional dark states except those of types-(a), (b), (c), (d) identified in \cite{mares2020}, which span the $\mathcal{K}$ and $\mathcal{M}$ subspaces. We denote the set of vertices connecting to the sink as $\delta V$, i.e. in our case we find $\delta V = \{2L, 2L-1\}$. The probability transition matrix with the Dirichlet boundary condition on $\delta V$ is denoted by$(2L+1)\times (2L+1)$ matrix $T$: for $j=0,\ldots 2L$, 
\begin{eqnarray}
\nonumber T_{i,j} & = & \frac{1}{3}\left(\delta_{j,i+1} + \delta_{j,i-2} + \delta_{j,i+2}\right), \quad i = 2, 3, \ldots , 2L-2, \\
\nonumber T_{i,j} & = & \frac{1}{3}\left(\delta_{j,i-1} + \delta_{j,i-2} + \delta_{j,i+2}\right), \quad i = 3, 5, \ldots , 2L-3, \\
\nonumber T_{0,j} & = & \frac{1}{3}\left(\delta_{j,0} + \delta_{j,1} + \delta_{j,2}\right), \quad T_{1,j} = \frac{1}{3}\left(\delta_{j,1} + \delta_{j,0} + \delta_{j,3}\right), \\
\nonumber T_{2L-1,j} & = & \frac{1}{3}\left(\delta_{j,2L-2} + \delta_{j,2L-3}\right), \quad  T_{2L,j} = \frac{1}{3}\left(\delta_{j,2L} + \delta_{j,2L-2}\right).
\end{eqnarray}
According to \cite{HiguchiSegawa,KonnoSegawaStefanak}, such an additional dark state has to belong into a subspace
    \begin{equation}
    \label{eq:RW} 
    \mathcal{T}= \bigoplus_{|\lambda|=1} \{ (1-\lambda R)d_1^*f \;|\; f\in   \ker((\lambda+\lambda^{-1})/2-T),\;\mathrm{supp}(f)\subset V\setminus \delta V\}.
    \end{equation} 
Here $d_1^*:\mathbb{C}^{V\setminus \delta V}\to \sH $ is a boundary operator defined by 
$$
(d_1^*g)(e):=\braket{e|d_1^*|g} = \frac{\braket{h(e)|g}}{\sqrt{\mathrm{deg}(h(e))}} , 
$$ 
for any $\ket{g}\in \mathbb{C}^V$ and $e\in E$, where $h(e)$ is the head of the arc $e\in E$. It is sufficient to consider the case (4) with all loops, in this case $\mathrm{deg}(u)=3$ for any $u$ since the graph is 3-regular.
Then, to show the non-existence of additional dark states, it is enough to show that $T$ has no eigenvectors whose supports are included in $V\setminus \delta V$.  
We proceed by contradiction. Assume there is such an eigenvector $f\in \mathbb{C}^V$, i.e. the relation $Tf=\mu f$ holds for some $\mu\in(-1,1)$. Since $f$ is not supported at $\delta V$ it holds that
$$
f(2L) = f(2L-1) = 0 .
$$
This leads us to 
\begin{align*} 
    (Tf)(2L) &= \frac{1}{3}f(2L-2))=\mu f(2L)=0 ,\\
    (Tf)(2L-1) &= \frac{1}{3}f(2L-3)+\frac{1}{3}f(2L-2)) = \mu f(2L-1)  = 0,
\end{align*}
which implies $f(2L-2)=f(2L-3)=0$. 
In the same way, 
    \begin{align*} 
    (Tf)(2L-2)) &= \frac{1}{3}\left(f(2L-4))+f(2L-1)+f(2L)\right) =\mu f(2(L-1)) =0 ,\\
    (Tf)(2L-3) &= \frac{1}{3}\left(f(2L-5)+f(2L-4)+f(2L-1)\right)=\mu f(2L-3) =0,
    \end{align*}
which implies $f(2L-4)=f(2L-5)=0$.
Using it recursively, it holds that $f(2k)=f(2k-1)=0$ for any $k=1,\dots,L$, and $f(0)=0$.  Then we have $f=0$, which is a contradiction. 
We conclude that there are no additional dark eigenstates except types (a)--(d). Hence, the orthogonal projectors derived in Sections \ref{sec:3} and \ref{sec:4} describe the whole dark subspace of the Grover walk on the ladder graph.

\section*{References}

\bibliography{biblio}

\bibliographystyle{iopart-num}

\end{document}